\documentclass[9pt,twocolumn]{extarticle}
\usepackage{graphicx}
\graphicspath{{./}{D:\images}}
\usepackage{amsmath}
\usepackage{color}
\usepackage{comment}
\usepackage{geometry}
\usepackage[
backend=biber,
style=numeric,
sorting=ynt
]{biblatex}
\title{PharML.Bind: Pharmacologic Machine Learning for Protein-Ligand Interactions
}
\author{{\small Aaron D. Vose$^{\dagger}$, Jacob Balma$^{\dagger}$, Damon Farnsworth$^{\dagger}$, Kaylie Anderson$^{\dagger}$, and Yuri K. Peterson$^{\ddagger}$} \\ {\scriptsize $^{\dagger}$Cray, a Hewlett Packard Enterprise Company, $^{\ddagger}$College of Pharmacy, Medical University of South Carolina}}
\begin{document}
\twocolumn[
  \begin{@twocolumnfalse}
    \maketitle
    \begin{abstract}
Is it feasible to create an analysis paradigm that can analyze and then accurately and quickly predict known drugs from experimental data?
PharML.Bind is a machine learning toolkit which is able to accomplish this feat. Utilizing deep neural networks and big data, PharML.Bind correlates experimentally-derived drug affinities and protein-ligand X-ray structures to create novel predictions.
The utility of  PharML.Bind is in its application as a rapid, accurate, and robust prediction platform for discovery and personalized medicine.
This paper demonstrates that graph neural networks (GNNs)
can be trained to screen hundreds of thousands of compounds against thousands of targets in minutes, a vastly shorter time than previous approaches.
This manuscript presents results from training and testing using the entirety of BindingDB after cleaning; this includes a test set with 19,708 X-ray structures and 247,633 drugs, leading to 2,708,151 unique protein-ligand pairings.
PharML.Bind achieves a prodigious 98.3\% accuracy on this test set in under 25 minutes.
PharML.Bind is premised on the following key principles:
1) speed and a high enrichment factor per unit compute time, provided by high-quality training data combined with a novel GNN architecture and use of high-performance computing resources,
2) the ability to generalize to proteins and drugs outside of the training set, including those with unknown active sites, through the use of an active-site-agnostic GNN mapping,
and 3) the ability to be easily integrated as a component of increasingly-complex prediction and analysis pipelines.
PharML.Bind represents a timely and practical approach to leverage the power of machine learning to efficiently analyze and predict drug action on any practical scale and will provide utility in a variety of discovery and medical applications.\\

\vspace{\baselineskip}
    \end{abstract}
  \end{@twocolumnfalse}
]

\section{Introduction}
\begin{figure*}[ht]
    \centering
    \includegraphics[width=1.0\textwidth]{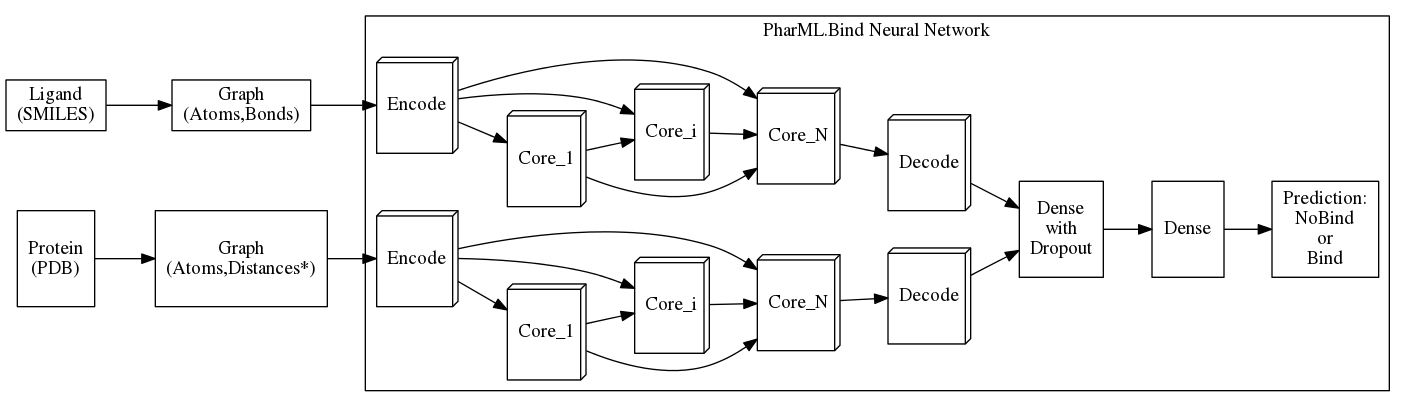}\\
    \vspace{0.75\baselineskip}
    \begin{minipage}{0.95\linewidth}
    \begin{small}
    \caption{Visual depiction of the PharML.Bind data inputs and neural network model, where GNN layers are represented as 3D blocks, and traditional TensorFlow layers are represented as 2D blocks. $^*$Distances in the protein graph are 1.0/distance with a minimum value of 1.0/4\AA; edges with a smaller value are pruned.}
    \label{fig:model}
    \end{small}
    \vspace{-2.0mm}
    \end{minipage}
\end{figure*}

ParmML.Bind is a novel, high-throughput, \textit{in silico} framework for the acceleration of critical portions of contemporary and future pharmaceutical drug-discovery workflows.
Specifically, PharML.Bind accelerates prediction of the binding of candidate drugs to target proteins through utilization of experimentally-derived 3D protein structures and associated drug affinities. This is achieved by applying state-of-the-art machine learning (ML) techniques to efficiently model the rules governing affinity in a data-driven manner.
Efficient and accurate drug discovery for target proteins, as well as discovery of relevant proteins for candidate drugs, is of primary interest to pharmaceutical companies, healthcare practitioners, and patients.

The area of computer-aided drug discovery (CADD) was born as computational performance advanced to the point that simulation of drug-target interactions was possible with reasonable speed and accuracy.
CADD was founded on the capabilities of computers to simulate drug-target interactions and proved to be an efficient and rational drug discovery tool.
However, in the last decade, the prominence of artificial intelligence (AI) and machine learning (ML) has been at the forefront of the latest and most-advanced drug discovery techniques.
Using AI as a prediction tool
is reaching peak hype, and for good reason, as there is enormous investment, excitement, and potential \cite{vamathevan2019applications} \cite{smith2019startups}.

Most software programs for drug docking predictions used to date depend upon approximations to estimate binding energy between the drug and target.
Among such programs are the popular Dock \cite{lang2009dock}, AutodockVina \cite{trott2010autodock}, GLIDE \cite{friesner2004glide}, and MOE \cite{moe2019}.
These programs stochastically rotate bonds at a pre-determined target binding site
and calculate the free energies of binding \textit{de novo} using derivations of the Schr\"{o}dinger equation.
As solving the Schr\"{o}dinger equation for a complex system has an enormous number of variables for which to solve, docking software was developed using density functional theory and analogy to Newtonian molecular mechanics.
This method simplifies binding by looking at the interaction as densities rather than electronic probability fields.
Such simplifications led to three major limitations in docking:
1) use of molecular mechanics versus molecular dynamics,
2) implicit versus explicit water,
and 3) the expectation of a linear response versus real world sigmoidal response.
However, recent developments in efficient software design and fast processing hardware means it is now feasible to correct these shortcomings.
Disparate research groups are addressing all of these problems using increasingly powerful accelerators such as graphics-processing units (GPUs) and application-specific integrated circuits (ASICs).

The shortcoming of these softwares are their expectation of a normal statistical response, when in actuality the response is inherently nonlinear and often stochastic.
This was done due to the large scaling in complexity of using non-linear math.
The complexity arises from interactions featuring noncovalent protein-ligand interactions which are a mixture of mostly hydrogen bonds, ionic bonds, and hydrophobic bonds.
Understanding and simulation of hydrogen bonds and ionic bonds are very good.
However, hydrophobic bond calculations are woefully underdeveloped.
The real world energy contributions of hydrophobic bonds is mostly derived by the exclusion of water and altering the quantum network of the adjacent hydration sphere.
However, current systems approximate this by taking two essentially \textit{in vacuo} phase molecules, and approximating the partial charges based on simplistic tables of the ionization of amino acids at a given hydrated pH.
This means all the quantum effects and much of the induced effects are not considered.
The reason this is not done currently is the difficulty in calculating entropy combined with the added complexity of docking being a closed bimolecular system.

However, artificial intelligence (AI) utilizing deep graph neural networks (GNNs) presents an entirely different way to solve the problem \cite{deepmindgraphnets}.
Rather than using rules based on approximations of physics or clustering similar molecules by organic chemistry / electronic properties, GNNs represent a form of processing by which learning is performed without the introduction of a large class of potentially-biased, human-created rules.
Instead of trying to derive the energy of a complex \textit{de novo}, ML can predict the probability of a complex forming though neural network pattern recognition.
The approach has been implemented on modest training scales in programs such as Quantum.Ligand.Dock \cite{kantardjiev2012quantum}, AlgoGen \cite{barberot2014algogen}, QPED \cite{kalyaanamoorthy2013quantum}, and PotentialNet \cite{feinberg2018potentialnet}.

Protein-drug interactions are the basis for a drug's mechanism of action, referred to as pharmacodynamics when speaking of the patient.
While it is possible to get direct experimental evidence of the specific interactions though techniques like X-ray and NMR spectroscopy, these techniques are arduous.
Predicting protein-ligand interactions is therefore of specific interest to scientists in academia and the pharmaceutical industry who toil in the drug discovery and development space.
Prediction of these interactions could help elucidate what the mechanism of action is for a current drug, predict the other proteins a ligand might bind to (i.e., off-target interactions), and predict
new classes of drugs for drug development campaigns.
In this regard, it can be argued the great need is not to create algorithms that best match \textit{in vitro} data quantitatively, but rather have the ability to accurately predict the qualitative ability of entities to interact in order to best prioritize quantitative real-world experiments (i.e., quality of predictions in some instances is more important than precision).

Most molecular interaction prediction software to date comes in two basic types: 1) docking with interaction energy equitation solving algorithms and 2) chemical subgraph or e-state comparison such as nearest neighbor and QSAR algorithms \cite{itskowitz2005k} \cite{peterson2009discovery}.
However, AI and ML approaches are gaining in number and interest but come in a much wider array of basic premises.
The premise from the inception of PharML.Bind was to do something substantially different from current paradigms, and in particular, to leverage the power of neural networks (NNs) to find answers, that from an empirical and mathematical point of view, could be very difficult to solve.
For instance, trying to solve the full Schr\"{o}dinger equation for a dimer in a full biological context is an overwhelming task.  The primary reason for this is that accurately stating the initial conditions is imprecise and complicated further still by the local minima problem, which is solved in docking by doing repeated measures.
Switching the paradigm from trying to accurately predict the true affinity, to being able to discriminate and rank probable from improbable, provides a robustness that is much less sensitive to small changes in initial conditions.

\section{Motivation and Related Work}\label{sect:motivation}
\begin{figure*}[!ht]
\centering
\begin{minipage}{0.33\linewidth}
    \centering
    \hfill
    \includegraphics[width=.8\textwidth]{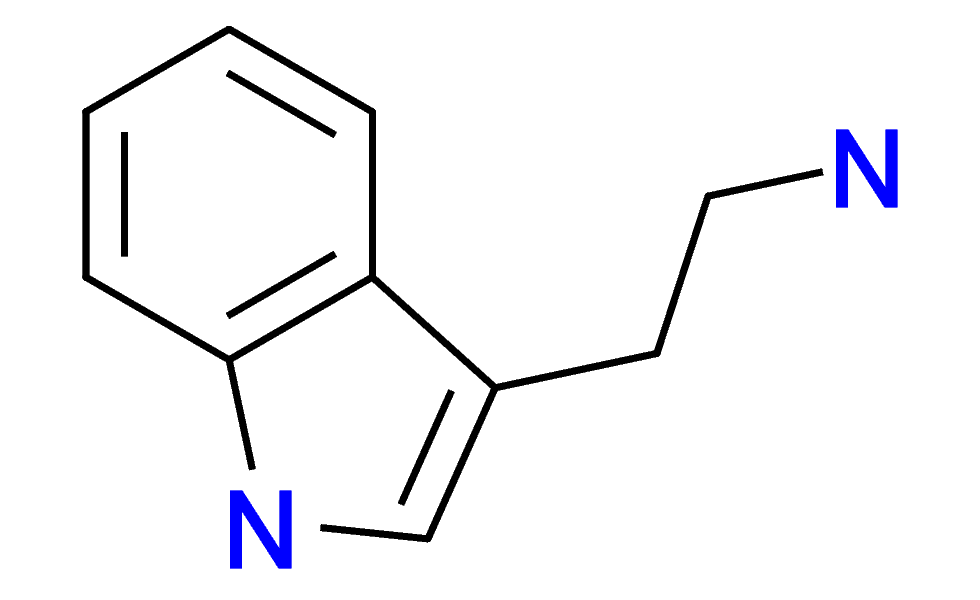}
\end{minipage}\hfill
\begin{minipage}{0.66\linewidth}
    \centering
    \begin{small}
    \begin{tabular}{lclc|lclc}
    \hline
    \multicolumn{4}{c|}{Edge$_i$ (Bond)} & \multicolumn{4}{c}{Node$_i$ \footnotesize{(Charge)}} \\
    \multicolumn{4}{c|}{\footnotesize{$(N_a,N_b,\textrm{BondType})$}} & \multicolumn{4}{c}{\footnotesize{(Atomic, Formal)}} \\
    \hline
    $E_1$ & (1,2,4) & $E_8$    & (8,4,4)   & $N_1$ & (6,0) & $N_8$ & (6,0)\\
    $E_2$ & (2,3,4) & $E_9$    & (8,9,4)   & $N_2$ & (6,0) & $N_9$ & (6,0)\\
    $E_3$ & (3,4,4) & $E_{10}$ & (9,1,4)   & $N_3$ & (6,0) & $N_{10}$ & (6,0)\\
    $E_4$ & (4,5,1) & $E_{11}$ & (5,10,1)  & $N_4$ & (6,0) & $N_{11}$ & (6,0)\\
    $E_5$ & (5,6,2) & $E_{12}$ & (10,11,1) & $N_5$ & (6,0) & $N_{12}$ & (7,0)\\
    $E_6$ & (6,7,1) & $E_{13}$ & (11,12,1) & $N_6$ & (6,0) & \\
    $E_7$ & (7,8,1) &          &           & $N_7$ & (7,0) & \\
    \hline
    \end{tabular}
    \end{small}
\end{minipage}\\
\vspace{0.75\baselineskip}
\begin{minipage}{0.95\linewidth}
\begin{small}
\caption{Two-dimensional chemical structure of tryptamine (left) along with the the corresponding undirected ligand graph as a list of nodes and edges as input into the PharML.Bind neural network (right). Both forms contain essentially the same information as the SMILES string ``\texttt{c1ccc2c(c1)c(c[nH]2)CCN}''.}
\label{fig:tryptamine}
\end{small}
\vspace{-2.0mm}
\end{minipage}
\end{figure*}

A major driver in drug discovery is the rate at which new drugs can be characterized relative to a given target.
It is helpful to define a metric to serve the purpose of characterizing the number of simulations per second one can utilize to rank protein-ligand pairings.
This work quantifies performance in terms of the number of protein-ligand pairings which can be classified as either Bind or No-Bind per second, that is protein-ligand pairs per second (PLP/s).

A 2017 review of various approaches to docking found significant variability in performance of predicted binding affinity accuracy across ten major docking applications \cite{docking_review2017}.
Although some tools exist to discriminate with good accuracy between different ligands based on binding affinity for specific targets, they rely on simulation or random search to make predictions, making them time-consuming to utilize in practice.
These tools usually provide a score of the ligand's pose relative to the target in 3D space or predict affinity directly based on predicted hydrogen bonding patterns.  In this case, \textit{in silico} accuracy is assessed in a non-data driven manner through multiple measures with the premise that correct answers will converge.

In contrast, training of NNs is data-driven, meaning that NNs learn a mapping function (i.e., input to output) from a distribution of examples, making them highly sensitive to the quality and quantity of training data.
It is generally believed that these models make use of hierarchies of learned features to make predictions.
However, unlike many other NNs, GNNs can learn to detect such features in the input data in a way which is more tolerant of many classes of input transformations.

As an example, rotation, translation, and other ordering transformations applied to input data used with contemporary convolutional neural networks (CNNs) and other common architectures can cause important features to be missed as learned features no longer line up with the transformed input data.
In order to learn rotational invariance or other such symmetries, a CNN can be trained with (potentially all) valid transformations applied to the input data, or can internalize some of these transformations \cite{dieleman2015rotation}.
However, this data augmentation approach can require training with up to $N!$ permutations of the data to make sure all possible orientations are covered during training, where $N$ is the number of input values fed to dense and similar layers in the neural network \cite{gens2014deep}.
This frequently represents an intractable amount of additional training time for large molecules and even small proteins.
In contrast, GNNs are able to process input data in a way which is invariant to rotation, translation, and other ordering transformations.
This effectively allows GNNs to learn tasks with fewer training examples.
Specifically, GNNs can learn the representation of a molecule and its relationship to a given protein target from only a single disjoint graph containing the two entities. 
This could otherwise require $N!$ permutations of the input data in order to be learned by a CNN or fully-connected network.
\vspace{-2.0mm}

\section{Methods and Key Results}\label{sect:methods_kr}
\begin{figure*}
    \centering
    \begin{small}
    \begin{tabular}{ccc}
    \includegraphics[height=5cm]{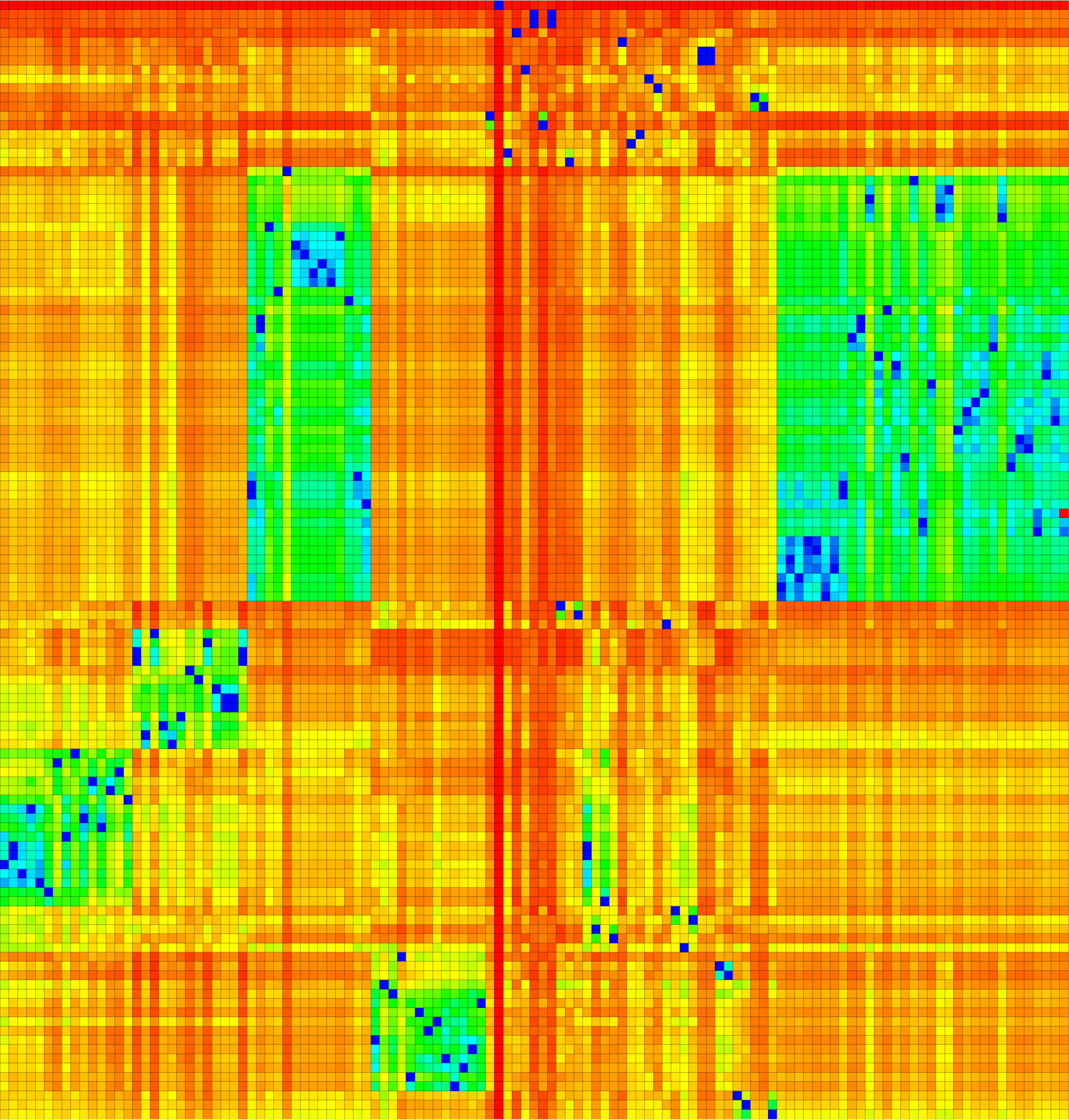} & $\:\:\:\:\:\:\:\:\:\:\:\:\:\:\:\:$ & \includegraphics[height=5cm]{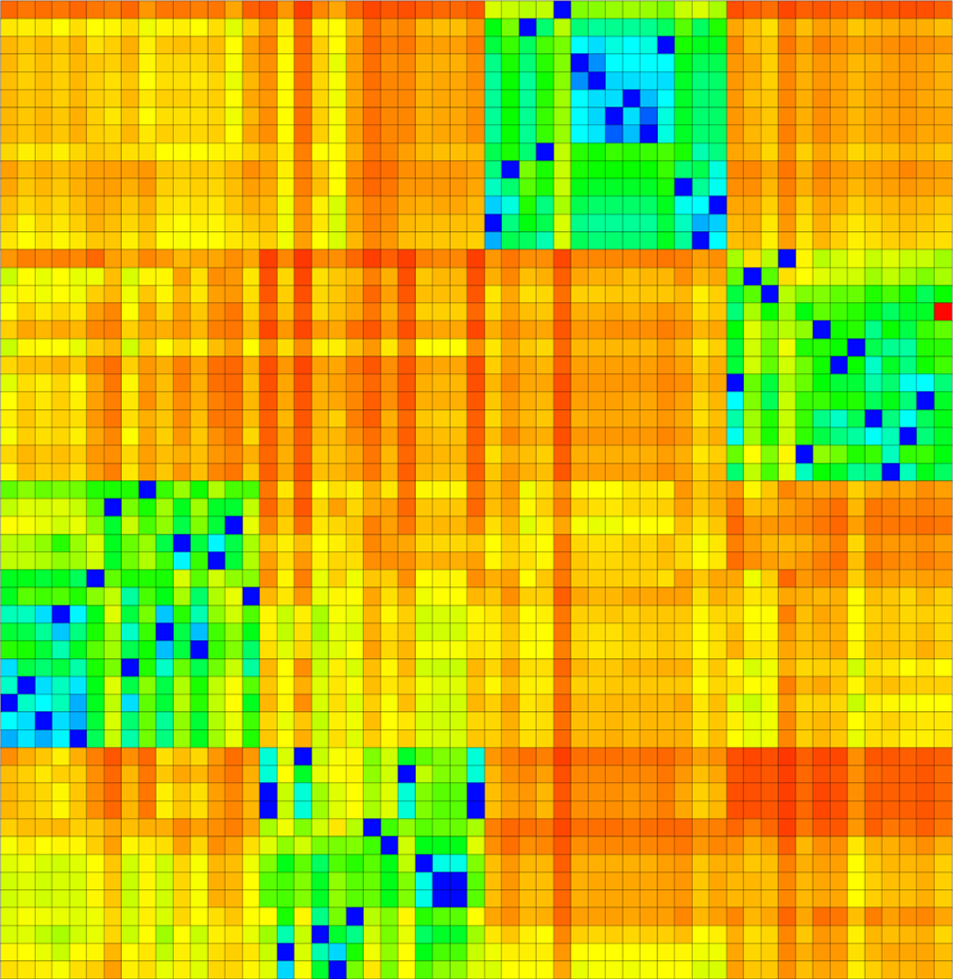} \\
    & & \\
    (a) Bind and No-Bind & & (b) Bind \\
    \end{tabular}
    \end{small}\\
    \vspace{0.75\baselineskip}
    \begin{minipage}{0.95\linewidth}
    \begin{small}
    \caption{Heat map pair-wise similarity matrices, indicating global diversity and local similarity of drugs. (a) Similarity of Bind and No-Bind ligands from the validation dataset (see Table \ref{tab:datasets}) across a variety of specific protein classes, as well as (b) similarity of data classified Bind from (a). Heat maps are generated using Chemmine \cite{backman2011chemmine}. Green is highly similar and red is highly dissimilar.\label{fig:sim_bn_b}}
    \end{small}
    \end{minipage}
\end{figure*}

This work's recommended approach uses an ensemble of GNNs which takes as input graph representations of protein-ligand pairings, and then outputs a binary classification, either Bind or No-Bind.
Dataset processing, model implementation specifics, and key results are provided in this section.

\subsection{Dataset Preparation and Analysis}
In molecular docking, using a data-driven approach (e.g., data coming from X-Ray, Cryo-EM, or NMR experiments) enables a model to utilize the physical properties of binding states which occur in experimental data.
The physical properties which the model can learn and use to make predictions are mainly limited by the resolution and experimental quality.
Using graphs for both the protein and ligand in a sparse format consisting of a list of edges and a list of nodes allows for more efficient use of processing and memory during training than is possible with rasterized and other such dense approaches.

A single input which is fed to the GNN is given as a protein-ligand pair.
To give a specific example of the ligand graph input format, consider the input graph for tryptamine, presented in Figure \ref{fig:tryptamine}, with hydrogen excluded for clarity.
Note that both representations in the figure contain the same information as the SMILES string ``\texttt{c1ccc2c(c1)c(c[nH]2)CCN}''.
On the protein side, the conversion from a PDB to a protein neighborhood graph (NHG) is similar, but uses different node and edge attributes.
Specifically, the protein NHG is the same as the ligand graph, except that the bond type is replaced with 1/distance in Angstroms (with a minimum value of 1/4\AA, else the edge is removed) and only the atomic charge is used with the formal charge excluded.

BindingDB (BDB) is processed from a single SDF data file, \begin{small}\verb'BindingDB-All-terse-2D-2019m7.sdf'\end{small},
to extract information on proteins, ligands, and their interactions \cite{binding_db_pmid11812264}\cite{binding_db_pmid11836221}\cite{binding_db_pmid11987162}.
Proteins in BindingDB are specified by Protein Data Bank (PDB) \cite{Berman00theprotein} IDs, ligands are specified by SMILES strings, and their associated affinities are in the form of IC$_{50}$ values.
These entries from BindingDB and PDB are then examined for issues, such as an invalid SMILES string (as interpreted by RDKit \cite{rdkit}) or a protein with too many ($>\!10,000$) or too few ($<500$) atoms, discarding items failing validation.
The resulting protein-ligand pairs are then classified as either Bind (IC$_{50}\!\leq$ 10,000 nM) or NoBind ($>$10,000 nM).
Finally, the PDB and SMILES data are converted to their respective graph formats.

Data produced by the cleaning and format conversion process is then balanced to contain an equal number of Bind and NoBind examples, after which it is split into training, validation, and test sets as listed in Table \ref{tab:datasets}.
This data splitting is performed on the protein axis, such that a given PDB ID occurs in exactly one dataset.
Notice that the largest training set used is only 25\% of the total available data from BindingDB after pre-processing.
The remaining 75\% of the dataset is used to test model generalization.

\begin{table}[ht]
\centering
\begin{small}
\begin{tabular}{l|rrrr}
\hline 
Dataset & BDB & Pairs & Proteins & Ligands \\
\hline
ensemble$_a$ & ~5\% & 167,656 & 1,293 & 85,591 \\
ensemble$_b$ & ~5\% & 155,567 & 1,283 & 82,646 \\
ensemble$_c$ & ~5\% & 163,288 & 1,294 & 85,441 \\
ensemble$_d$ & ~5\% & 161,341 & 1,291 & 90,238 \\
ensemble$_e$ & ~5\% & 174,342 & 1,291 & 87,604 \\
\hline
mono$_{10}$ & ~10\% & 323,223 & 2,576 & 121,871 \\
mono$_{15}$ & ~15\% & 486,511 & 3,870 & 147,632 \\
mono$_{20}$ & ~20\% & 647,852 & 5,161 & 167,086 \\
mono$_{25}$ & ~25\% & 842,693 & 6,496 & 183,126 \\
\hline
validation & ~1\%  & 26,051 & 259 & 22,286 \\
testing    & ~75\% & 2,708,151 & 19,709 & 247,633 \\
\hline
\end{tabular}\\
\vspace{0.75\baselineskip}
\begin{minipage}{0.95\linewidth}
\caption{Details for ensemble and monolithic training datasets, as well as datasets used for validation to detect convergence. Details on the test set used for final evaluation are also provided.}
\label{tab:datasets}
\end{minipage}
\vspace{-3.0mm}
\end{small}
\end{table}

The threshold IC$_{50}$ score which was used to categorize pairings as Bind or No-Bind can be analyzed for fairness through similarity clustering analysis, as shown in Figure \ref{fig:sim_bn_b}a along with the similarity among pairs within only the Bind category is shown in Figure \ref{fig:sim_bn_b}b.
These data indicate splitting the data would have equivalent effects on training for both Bind and No-Bind, leading to  more robust prediction for either actives or inactives. Very uneven splits would bias the predictions to have accuracy for \textit{either} Bind or No-Bind.

\subsection{Model Architecture and Training}\label{sec:arch-imp-train}
The choice of model architecture has important ramifications for allowable mappings (i.e., inputs to outputs), achievable accuracy, and generalization capabilities.
Thus, the GNN architecture created for PharML.Bind is designed to have an active-site-agnostic mapping and architectural details which optimize training and inference with graph-based chemical data.
The model is implemented in Python using TensorFlow \cite{tensorflow2015-whitepaper}.

CNN architectures commonly used with image data, such as architectures similar to those used by LeNet \cite{lecun1998gradient} and ResNet \cite{he2016deep}, do not directly propagate information from earlier layers all the way to the later dense layer(s). 
Initial features extracted by the first convolutional layers are simple gradients and basic curves, while what is most important for input to the final classification layers are the higher-level, more-complex features extracted by the last convolutional layer.
However, compared to traditional CNNs on image data, even low-level features involving only a few atoms can be very important in the chemical space.
For this reason, the PharML.Bind NN architecture carries all output features forward as each message-passing layer is added, as can be seen in Figure \ref{fig:model} and Table \ref{tab:topo}.
Also, note that the sizes of the MLP NNs used for $\phi$ in the cores increase with depth, as later layers have more input features to consider than earlier layers.

Weights for each message-passing step (i.e., Core$_i$) are independent and not shared with other cores.
This design choice is motivated by the idea that each Core$_i$ transforms one input latent representation $L_i$  into another output latent representation $L_j$, and it is undesirable to ``overload'' the NNs which make up each core.
That is, each core / message-passing step should have its own $L_i \rightarrow L_j$ mapping, as opposed to other common approaches that give each \textit{edge type} independent weights but share weights across steps \cite{gilmer2017neural}.

\begin{table}[!ht]
    \centering
    \begin{small}
    \begin{tabular}{l|l}
    \hline
    GNN Layer & Layer Sizes for MLP $\phi$ \\
    \hline
    Encode & 32, 32 \\
    Core$_0$ & 32, 7 \\
    Core$_1$ & 40, 9 \\
    Core$_2$ & 48, 29, 48 \\
    Core$_3$ & 56, 41, 27, 56 \\
    Core$_4$ & 64, 51, 39, 51, 64 \\
    Decode & 32, 32 \\
    Output & 64 \\
    \hline
    TF Layer & Size \\
    \hline
    Dense$_0$ & 128 with dropout 0.5\\
    Dense$_1$ & 128 \\
    Output & 2 \\
    \hline
    \end{tabular}\\
    \vspace{0.75\baselineskip}
    \begin{minipage}{0.95\linewidth}
    \caption{Layer sizes used, where $\phi$ is a multilayer perceptron (MLP) for each of $\phi^e$, $\phi^v$, and $\phi^u$ in PharML.Bind's Core$_i$ layers as well as the Encode and Decode GNN layers \cite{deepmindgraphnets}.  The output GNN layer only uses $\phi^u$.}
    \label{tab:topo}
    \end{minipage}
    \vspace{-4.0mm}
    \end{small}
\end{table}

A cyclical learning rate (CLR) was used during training with a triangular waveform which completed one cycle every 5 epochs \cite{itskowitz2005k}.
This use of CLR was combined with a learning rate decay schedule by setting the base learning rate used by the CLR during training to the decay function.
An initial learning rate of {$\ell_0$=$1\!\times\!10^{-8}$} was used, and scaled by {$N_{workers}$} as additional workers were added to speed up the training time.
Convergence was defined as 15 training epochs without improvement in validation accuracy. A 1\% subset of the 75\% test set was used to test for convergence.
This was found to be a decent proxy for performance on the full 75\% test, and allowed for a validation phase to be run after each epoch to check for convergence without impacting the overall training time significantly. 

\subsection{Experimental Results}\label{sec:exp_results}
Results from performed experiments demonstrate that test accuracy and generalization improve as more training data are used.
For the largest dataset, 25\% of the available pairings in BindingDB were used for training, with 75\% for testing.
This is an inversion of the size ratio utilized for traditional training and test datasets, but quite useful to illustrate that the Bind / No-Bind prediction task has truly been learned by the model in a way that generalizes across many more pairings than ever observed during training.
This result implies the model has learned actionable physical rules from the data given.
Overall, the best model was able to achieve 98.3\% accuracy on the largest test set, containing 2,708,151 pairs from BindingDB in under 25 minutes.
This represents use of 100\% of the full, filtered BindingDB dataset, the model being trained on only 25\% of it, with the remainder reserved for validation and testing.

\begin{figure}[!ht]
\centering
  \includegraphics[width=1.0\linewidth]{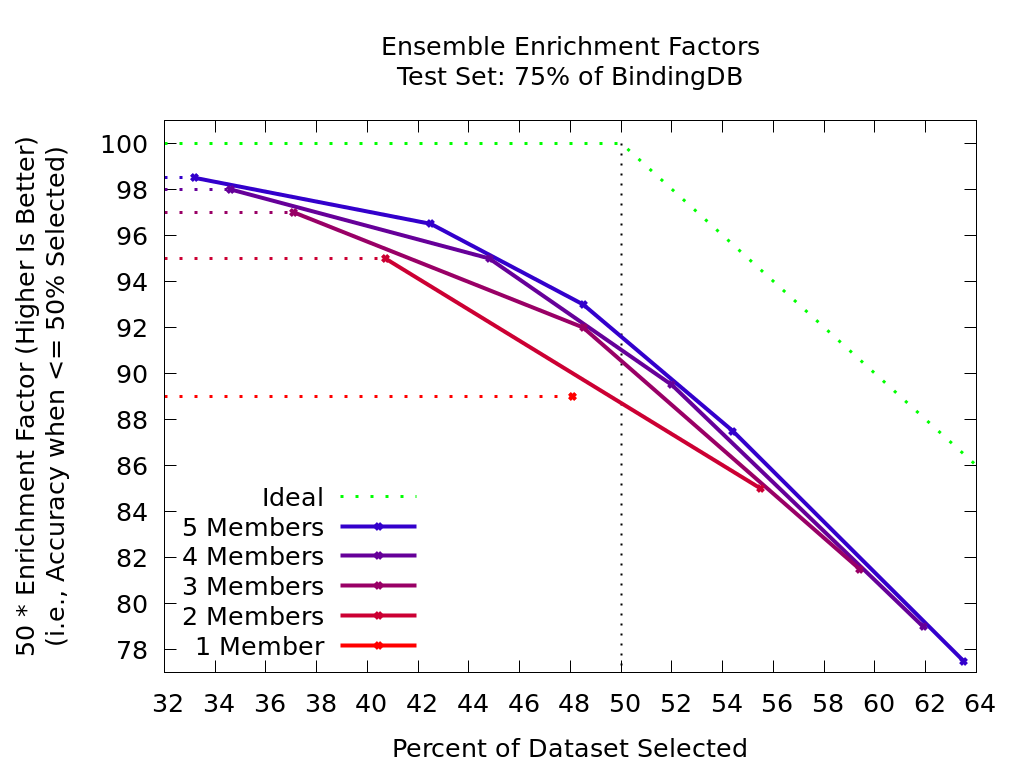}
\vspace{0.75\baselineskip}
\begin{minipage}{0.95\linewidth}
    \caption{Enrichment factors \cite{pearlman2001improved} and accuracy for PharML.Bind testing with different ensemble sizes on BindingDB.\label{fig:acc_ensmbl_mono}}
    \vspace{-2.0mm}
\end{minipage}
\end{figure}

\begin{figure}[ht]
\centering
\includegraphics[width=1.0\linewidth]{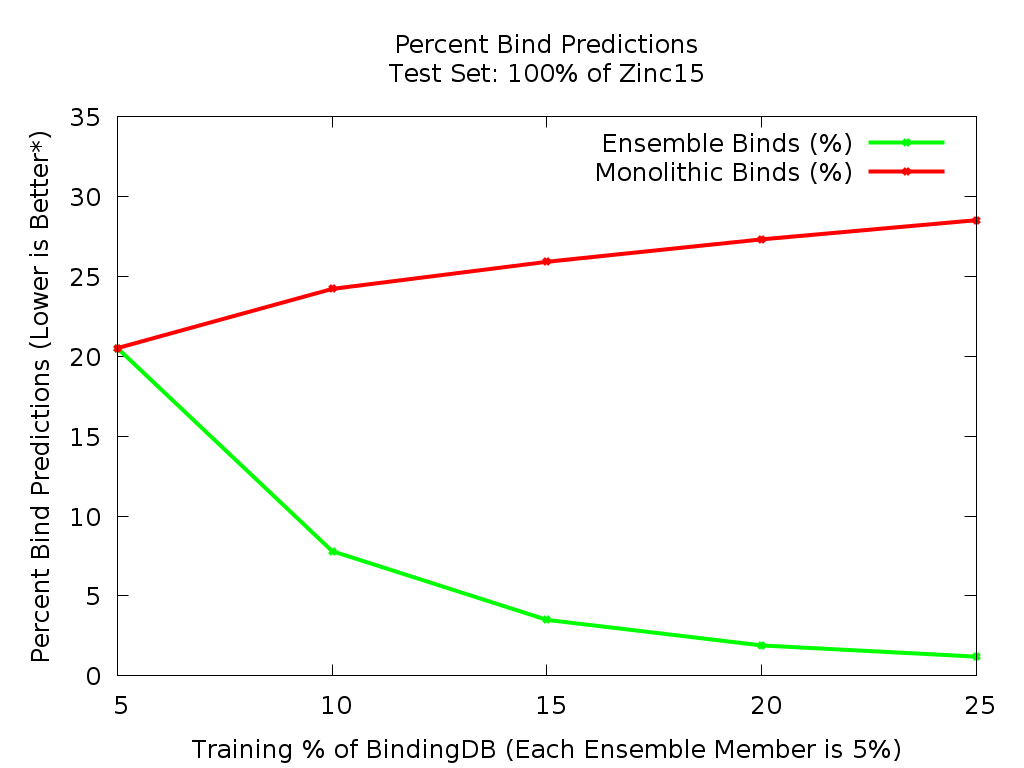}
\begin{minipage}{0.95\linewidth}
\begin{small}
\caption{Percent of the Zinc15 dataset predicted to bind to the target protein for varying numbers of ensemble members as well as monolithic NNs.\label{fig:acc_zinc}}
\end{small}
\end{minipage}
\end{figure}

Table \ref{tab:res-bdb} provides results which show that an ensemble of multiple GNNs provides a higher accuracy than training a single monolithic GNN with the same amount of data.
These results from testing with BindingDB alone may not adequately motivate the use of ensembles due to the requisite increase in model complexity as well as the associated increase in training and inference time.
However, results from testing performed with the ZINC15 dataset \cite{sterling2015zinc} against an instance of the 5-HT receptor (i.e., PDB ID ``5TVN'') as presented in Figure \ref{fig:acc_zinc} do provide solid motivation for the use of ensembles.
As shown in the figure, monolithic GNNs trained with increasing dataset sizes result in an increase in predicted binds from ZINC15, with the largest training set producing a bind prediction for more than 25\% of the ligands (a similar increasing trend is seen with BindingDB in Table \ref{tab:res-bdb}). This bind prediction rate is too high to be considered accurate.
In contrast, the ensemble approach predicts binds using ZINC15 at a much more realistic and useful rate of approximately 1.2\%, while at the same time performing better on BindingDB.

\begin{table}[!ht]
  \centering
  \begin{small}
  \begin{tabular}{r|cc}
      \hline
      Train & Accuracy (\%) & Bind Predictions (\%) \\
      (BDB\%) & \multicolumn{2}{c}{(Ensemble/Monolithic)} \\
      \hline
      5  & 88.9 / 88.9 & 48.1 / 48.1 \\
      10 & 95.2 / 93.2 & 40.7 / 48.3 \\
      15 & 97.0 / 95.1 & 37.1 / 48.8 \\
      20 & 97.9 / 96.2 & 34.6 / 49.2 \\
      25 & 98.3 / 96.8 & 33.2 / 49.7 \\
      \hline
  \end{tabular}\\
    \vspace{0.75\baselineskip}
    \begin{minipage}{0.95\linewidth}
    \caption{Accuracy of monolithic training runs compared to ensembles (each ensemble member trains on 5\% of BindingDB).}
    \label{tab:res-bdb}
    \end{minipage}
    \vspace{-5.0mm}
    \end{small}
\end{table}

The effectiveness of the ensemble approach might be explained by the dissimilarity between the ligands found in BindingDB and ZINC15.
A GNN will be more likely to make an essentially random prediction for a ligand unlike those it has been trained with as compared to those ligands it has seen during training.
Thus, an ensemble of GNNs will be able to probabilistically reject these ``uncertain'' predictions at a higher rate than the monolithic approach.

\subsection{Performance}
As described in Section \ref{sec:exp_results}, GNNs were used to inference the test set after training converged.
The test set was run on a cluster consisting of Cray CS-Storm servers with 4 V100 GPUs attached to each socket of Intel E52697-v4 36-core host processors.
PharML.Bind achieved a record-breaking rate of 195 protein-ligand pairs per second (PLP/s) on one state-of-the-art Nvidia V100 GPU. When using 16 servers with 64 V100s, performance scales to over 3000 PLP/s.
Speed improvement is given by:

\begin{equation}
    R_{plp} = \alpha K_{gpu} \frac{N_{plp}}{\Delta t},
\end{equation}

\noindent where $R_{plp}$ is the rate of protein-ligand pairs inferenced per second (PLP/s), $\alpha$ is a scaling factor, $K_{gpu}$ is the number of GPU resources utilized, $N_{plp}$ is the number of protein-ligand pairs inferenced over the time interval $\Delta t$, and $\Delta t$ is the time spent inferencing $N_{plp}$ examples.

As a proxy for and instance of traditional simulation methods, MOE Dock was used to predict affinity for 83 compounds across the whole surface of a CA4 target protein.
The simulation was run for ~105 hours on a 4-core Xeon W3565 3.2GHz processor. Over the course of the simulation, all 8 threads of the CPU were utilized 75\% of the time.
We use this performance as a baseline model of MOE Dock to estimate how the application would theoretically perform when run on the same host CPU hardware comprising the servers on which PharML.Bind was run. 

MOE Dock achieves $2.19\! \times\!10^{-4}$ PLP/s on a Xeon W3565.
Adjusting for differences between the systems, MOE Dock is projected to achieve $136.2\! \times\!10^{-4}$ PLP/s on an extrapolated system to that used with PharML.Bind.
Using the performance model of MOE described above, PharML.Bind offers a speedup factor of over 14,000x improvement over MOE Dock in terms of raw time to arrive at actionable predictions about protein-ligand pairings.
We can bound the improvement factor by setting the MOE performance model to be either 100\% compute-bound, or 100\% memory bandwidth bound, which assumes that MOE Dock performance is boosted entirely by either the improvements in memory bandwidth or compute capability of the E52697-v4 36-core server relative to the Xeon W3565 based server.
These bounds result in a minimum speedup of 11,000$\!\times$ and a maximum speedup of 37,000$\!\times$, as shown in Table \ref{tab:performance}.

\begin{table}[!ht]
    \centering
    \begin{small}
    \begin{tabular}{l|r}
    \hline
    System Model & PLP/s  \\
    \hline
    PharML.Bind on 4 GPUs & 195 \\
    MOE on 8 CPUs & 136.2$\times\!10^{-4}$ \\
    Upper Bound (Mem. BW Limited) & 52.6$\times\!10^{-4}$ \\
    Lower Bound (Compute Limited) & 164.1$\times\!10^{-4}$ \\
    \hline
    System Model & PharML.Bind \\
                 & Speedup \\
    \hline
    MOE on 8 CPUs & 14,363$\!\times$ \\
    Upper Bound (Mem. BW Limited) & 37,215$\!\times$ \\
    Lower Bound (Compute Limited) & 11,922$\!\times$ \\
    \hline
    \end{tabular}\\
    \vspace{0.75\baselineskip}
    \begin{minipage}{0.95\linewidth}
    \caption{Performance comparison showing PharML.Bind speedup factor over MOE Dock along with upper and lower bounds implied by the performance model.}
    \label{tab:performance}
    \end{minipage}
    \vspace{-2.0mm}
    \end{small}
\end{table}

\section{Summary}\label{sect:conclusion}
By utilizing state-of-the-art deep graph networks, PharML.Bind provides modular building blocks to enable high-throughput drug discovery.
This data-driven approach enables early affinity-based decisions on drug viability for specific whole-protein targets. The efficient performance of the distributed training and inference algorithms can scale linearly with compute resources if provided sufficient volumes of data. 
This work demonstrates that increasing the amount of training data fed to a particular network improves its generalization capabilities.
By breaking the training data into non-overlapping chunks, ensembles of smaller models can be trained. 
This work demonstrates that the aggregate performance of an ensemble against a single monolithic model trained on the same fraction of the data can significantly improve generalization and simultaneously provide mechanisms for drug scoring. 
Trained GNN models have the potential to replace traditional protein-drug docking approaches (e.g., simulation) if such ML approaches can generalize sufficiently well to previously unseen drug-target pairings. 

In the frequent case that a desired end goal is to determine if a drug is a viable candidate for further investigation relative to a whole protein, PharML.Bind demonstrates the ability to predict affinity across 75\% of the protein-ligand pairings present in BindingDB with over 93\% accuracy using a single GNN.
Furthermore, this work shows that the accuracy of the model can be improved by increasing the amount of training data sufficiently (see Table \ref{tab:res-bdb}), making it interesting as a potential means of making efficient use of the ever-expanding databases of X-Ray, NMR, and CryoEM experiments.
When more data are used in training, that accuracy can jump to over 96\% in testing on 75\% of BindingDB with a single model. 
Utilizing ensembles of GNNs trained on subsets from the same 25\% of BindingDB, the accuracy on the 75\% test set exceeded 98\% and allowed the rank-ordering of the tested compounds.
These accuracies are expected to improve with hyperparameter optimization (HPO), as the layer sizes and training regime used in this work represent a very small, manual search of the available hyperparameter space.
Future work includes plans to use population-based training (PBT) \cite{vose2019recombination}\cite{jaderberg2017population}.

In addition, PharML.Bind makes use of scalable data-parallel training algorithms, meaning the time it takes to train a new model on even the largest databases can be scaled down linearly with increasing computational resources.
This same technique lends itself to support large-scale parallel inference. 
The model uses a graph-based approach, which aside being far more efficient than volumetric approaches also enables efficient learing of mappings from common features contained within PDB and SDF data structures to actionable results.
By utilizing graph neural networks and large HPC systems, drug discovery can be scaled to unprecedented levels in terms of both prediction accuracy and throughput. 

\subsection{Conclusions}
The potential impact from tools which predict affinity for arbitrary target proteins in a scalable manner opens up the possibility of finding candidate drugs across the entire corpus of experimental data gathered through CryoEM, NMR, and X-Ray crystalography data available to date.
This expanding frontier of available data enables solutions to previously intractable problems. Our results demonstrate that ensembles of PharML.Bind GNN models can be trained on small, independent subsets of BindingDB (or similar PDB / SDF based datasets) to provide rank-ordered candidate drug listings across tens-of-thousands of proteins and millions of drugs in an active-site-agnostic manner.
The resolution of the ranking can be improved simply by training more models. 

Furthermore, the number of nodes and edges that define the graphs used to represent the ligand and protein pairing is limited only by memory capacity, meaning that mappings against large-molecule or protein-protein pairings are likely amenable to similar speedup factors as those shown in this work.
This is especially critical for unannotated proteins which currently represent about a fifth of the human transciptome \cite{samandi2017deep}.

%

%
\subsection{Acknowledgments}
Corresponding Authors: petersy@musc.edu (Medical University of South Carolina) and avose@cray.com (Cray, a Hewlett Packard Enterprise company). The team would like to thank Richard Trager for his assistance during research on protein-ligand binding prediction with CNNs. Additionally, the authors would like to thank Alainna White, Benjamin Robbins, and Rangan Sukumar for their valuable consultations.
The authors would also like to extend their thanks to Cray for contributing compute time on their internal XC and CS supercomputer and cluster systems, respectively.
\section{References}
\begingroup
\renewcommand{\section}[2]{}%
\printbibliography
\endgroup

\end{document}